\documentstyle[aps,preprint,tighten,prd]{revtex}

\begin{document}
\draft
\preprint{\parbox{32mm}{\raggedleft WUE-ITP-96-005\\
  hep-ph/9604439\\[5mm]}} 
\title{Chargino-Sneutrino Production in Electron-Photon Collisions}
\author{S.~Hesselbach\thanks{email: hesselb@physik.uni-wuerzburg.de}
and H.~Fraas}
\address{Institut f\"ur Theoretische Physik, Universit\"at
W\"urzburg\\ 
         D-97074 W\"urzburg, Germany}
\date{September 1996}
\maketitle
\begin{abstract}
We study the production of charginos and sneutrinos in
electron-photon collisions ($e^-  \gamma \to \tilde{\chi}^-_j
\tilde{\nu}_e$) 
within the minimal supersymmetric standard model
(MSSM). The high energy photons can be generated by Compton
backscattering of intense laser pulses off one of the beams of the
Next Linear Collider (NLC).
This process could offer a significant opportunity to identify
sneutrinos, which exclusively decay invisibly into a
neutrino and the LSP, since 
the cross sections are two orders of magnitude higher than for
the radiative production  of invisible sneutrinos ($e^+ e^- \to
\tilde{\nu} \bar{\tilde{\nu}} \gamma$).
For three scenarios of gaugino-higgsino mixing the cross sections
and polarization asymmetries are computed and the resulting signatures
are compared with the SM background.
\end{abstract}

\narrowtext

\section{Introduction}

As a characteristic feature of supersymmetry (SUSY) with
R-parity conservation all superparticles decay into the stable
lightest supersymmetric particle (LSP) which is usually assumed to be 
the lightest neutralino $\tilde{\chi}^0_1$. Since the LSP is weakly
interacting, 
it escapes detection and carries missing energy ($\mbox{$\not\!\!E$}$)
as the most 
distinctive signature of SUSY events. It may, however, happen that the
sneutrino is lighter than both the second lightest neutralino and the
lightest chargino. In this case the only kinematically allowed
two-body decay 
of the sneutrino is the invisible channel $\tilde{\nu}_e \to \nu_e
\tilde{\chi}^0_1$. Light sneutrinos are allowed in supergravity models
with 
common scalar and gaugino masses at the unification scale \cite{datta}
and have been until now not experimentally excluded \cite{aleph,lep}.
Then the only way to identify sneutrino production in
$e^+e^-$-annihilation is the
radiative process $e^+e^- \to \tilde{\nu} \bar{\tilde{\nu}} \gamma$
\cite{franke}. In the energy region between 100 and 500~GeV, however,
the cross sections for this process are rather small, of the order
of $10^{-2}$~pb, so that due to the large irreducible background from
$e^+e^- \to \nu \bar{\nu} \gamma$ it would be very difficult to
identify an invisible sneutrino. Also the use of polarized beams
cannot substantially improve the situation \cite{franke}.
Recently Datta, Guchait and Drees \cite{datta} discussed in detail
the impact of scenarios with invisible sneutrinos on SUSY searches at
LEP 200, leading to dramatic effects on the signal for pair production
of light charginos.

We study in this paper an alternative option: the
associated production of light sneutrinos and charginos in
electron-photon collisions $e^-\gamma \to
\tilde{\nu}_e\tilde{\chi}^-_j$ within the 
scope of the minimal supersymmetric standard model (MSSM) (see,
e.g.,~\cite{haberk}). Since
radiative sneutrino pair production in $e^+e^-$-annihilation is a
higher order process, the $e^-\gamma$-option should lead to drastically
increased cross sections.
Beyond that one expects for a light and invisible
sneutrino a clear signature since in this case the light chargino
decays almost completely leptonically into a single lepton and a
neutrino. We give in the following the cross sections for $e^-\gamma
\to \tilde{\nu}_e\tilde{\chi}^-_j$ for photons produced by
backscattering of laser 
pulses on a high energy electron beam \cite{ginzburg,borden} and
confront scenarios with 
light invisible sneutrinos and heavy sneutrinos decaying
visibly. Since experiments with suitable polarized
beams may be helpful to discriminate between the supersymmetric
process and the standard model (SM) background \cite{cuypers2}, we
discuss the polarization asymmetries of the convoluted cross sections
for different polarizations of the electron beam and the laser
beam. For unpolarized and monochromatic photons this process has been
investigated by Grifols and Pascual \cite{grifols}. The authors
confined themselves, however, to pure wino-like charginos with masses
now experimentally excluded.

This paper is organized as follows:
In sec.~\ref{sec2} we give the analytic formulae for the cross sections
and polarization asymmetries in the electron-photon center of mass
system (CMS)
as well as in the laboratory system. 
After choosing three characteristic gaugino-higgsino scenarios we
present in sec.~\ref{sec3} the numerical results
for monochromatic photons and for the $e$-$\gamma$ mode of a
linear collider. 
Finally we discuss the supersymmetric signatures in comparison with
the SM background.

\newpage
\section{Cross sections and polarization asymmetries}
\label{sec2}

The Feynman graphs contributing to the process 
$ e^- + \gamma\longrightarrow\tilde{\chi}^-_j + \tilde{\nu}_e $
($j=1,2$) are shown in 
fig.~\ref{feyn1}. Denoting the four-momenta of $e^-$, $\gamma$,
$\tilde{\chi}^-_j$ and $\tilde{\nu}_e$ with $p$, $k$, $p'$ and $k'$,
we introduce the 
Mandelstam variables $s=(p+k)^2$, $t=(k-p')^2$ and $u=(p-p')^2$.
$\epsilon^\mu$ is the polarization vector of the photon, $S^\mu$
($S^{\prime\mu}$) the spin vector of the electron (chargino). The
relevant couplings of the supersymmetric particles can be deduced from
the following interaction Lagrangians of the
MSSM (as for notations and conventions, we
closely follow \cite{haberk}):
\begin{equation} 
{\cal L}_{e\tilde{\nu}_e\tilde{\chi}}  =  -g \sum_{j=1}^2  (V_{j1}^* 
\bar{\tilde{\chi}}_j^c P_L e\tilde{\nu}_e^*  + V_{j1} \bar{e} P_R
\tilde{\chi}_j^c\tilde{\nu}_e )   ,  \label{Lesnc}   
\end{equation}
\begin{equation} 
{\cal L}_{\tilde{\chi}\tilde{\chi}\gamma}  =  -e A_\mu \sum_{j=1}^2
 \bar{\tilde{\chi}_j}\gamma^\mu  \tilde{\chi}_j   .   \label{Lccg}  
\end{equation}
For completeness, we add the Lagrangian for the $ee\gamma$-coupling
\begin{equation} \label{Leeg}
{\cal L}_{ee\gamma} = e A_\mu \bar{e} \gamma^\mu e.
\end{equation}
In equs.~(\ref{Lesnc}) -- (\ref{Leeg}) $\tilde{\chi}_j$ ($j=1,2$) and
$e$ are 
the four-component spinors of the chargino and the electron, while
$\tilde{\nu}_e$ is the field of the
electron-sneutrino. $\tilde{\chi}_j^c$ are charge 
conjugated spinor fields. Furthermore, $P_{R,L} = (1\pm \gamma_5)/2$
denote the right- and left-handed projection operators,
$g=e/\sin\theta_W$ 
($e>0$), and $V_{ij}$ is the $2\times 2$ unitary matrix appearing in
the diagonalization of the wino-charged higgsino mass matrix (see
\cite{bartlfraascharg} for more details).

The differential cross section for circularly polarized photons with
helicity $\lambda_\gamma$ and electrons with helicity $\lambda_e = \pm
1$ is given by
\begin{equation} \label{dsigcirc}
\frac{d\sigma}{d\Omega} =  \frac{d\sigma^0}{d\Omega} (1-\lambda_e)
			   (1+\lambda_\gamma a_c).
\end{equation}
In the electron-photon CMS the cross section for unpolarized beams is
\widetext
\begin{equation}   \label{dsig0} 
\frac{d\sigma^0}{d\Omega} = 
  \frac{|V_{j1}|^2\alpha^2}{8 \sin^2\theta_W} \frac{w}{s^2} 
  \Bigg\{ 
  \frac{\displaystyle 2s+4\Delta_j\left(1+\frac{\Delta_j}{s}\right)}
	{s+\Delta_j-w\cos\theta} 
  	{}- \frac{8 m_{\tilde{\chi}^\pm_j}^2
   \Delta_j}{(s+\Delta_j-w\cos\theta)^2} 
   - \frac{3s+3\Delta_j+w\cos\theta}{2s}  \Bigg\} ,
\end{equation}
and the circular polarization asymmetry $a_c$ is given by
\begin{equation} \label{ac}
a_c \frac{d\sigma^0}{d\Omega}  = 
     \frac{|V_{j1}|^2\alpha^2}{8 \sin^2\theta_W} \frac{w}{s^2}  
     \Bigg\{ 
     \frac{\displaystyle 2s + 4\Delta_j + 4m_{\tilde{\chi}^\pm_j}^2}  
		{s+\Delta_j-w\cos\theta} 
  {}- \frac{8 s m_{\tilde{\chi}^\pm_j}^2}{(s+\Delta_j-w\cos\theta)^2} 
    - \frac{3s+3\Delta_j+w\cos\theta}{2s}  \Bigg\}, 
\end{equation}
where
\begin{equation}
 w  \equiv  \sqrt{s-(m_{\tilde{\chi}^\pm_j}+m_{\tilde{\nu}_e})^2} 
     \sqrt{s-(m_{\tilde{\chi}^\pm_j}-m_{\tilde{\nu}_e})^2}
\mbox{\quad and\quad}
 \Delta_j  \equiv  m_{\tilde{\chi}^\pm_j}^2 - m_{\tilde{\nu}_e}^2  
	\label{deltaj}.
\end{equation}
Similarly for the total cross section for circularly polarized photons
we obtain
\begin{equation}
\sigma = \sigma^0 (1-\lambda_e) (1+\lambda_\gamma A_c) \label{totwq}
\end{equation}
with the unpolarized cross section
\begin{equation}
 \sigma^0 = 
 \frac{|V_{j1}|^2 \alpha^2 \pi}{4 \sin^2\theta_W}\frac{w}{s^3} 
   \left\{ \frac{\displaystyle 2s(s + 2\Delta_j) +
4\Delta_j^2}{w} 
       \ln\frac{s+\Delta_j+w}{s+\Delta_j-w} -3s -7\Delta_j \right\}.
\end{equation}
The circular polarization asymmetry is given by
\begin{equation}
 A_c \sigma^0 = \frac{|V_{j1}|^2 \alpha^2
         \pi}{4\sin^2\theta_W}\frac{w}{s^3} 
  \left\{ \frac{2s(s +2\Delta_j
+2m_{\tilde{\chi}^\pm_j}^2)}{w} 
        \ln\frac{s+\Delta_j+w}{s+\Delta_j-w} -7s -3\Delta_j \right\}.
\end{equation}

\narrowtext
For linearly polarized photons the differential cross section shows an 
azimuthal $\cos 2\phi$ dependence:
\begin{equation}
 \frac{d\sigma}{d\Omega} =  \frac{d\sigma^0}{d\Omega} (1 + a_l \cos
2\phi) 
\end{equation}
with
\begin{eqnarray}
a_l \frac{d\sigma^0}{d\Omega} & = & \frac{1}{2} 
	  (\frac{d\sigma_\|}{d\Omega}-\frac{d\sigma_\perp}{d\Omega})
	= \frac{d\sigma_\|}{d\Omega} - \frac{d\sigma^0}{d\Omega}
	= \frac{d\sigma^0}{d\Omega} - \frac{d\sigma_\perp}{d\Omega}
\nonumber\\ 
 & = &  \frac{|V_{j1}|^2 \alpha^2}{4 \sin^2\theta_W} 
      \frac{w^3}{s^3} \frac{\Delta_j
\sin^2\theta}{(s+\Delta_j-w\cos\theta)^2}. 
 \label{al}
\end{eqnarray}
In equ.~(\ref{al}) $\frac{d\sigma_\|}{d\Omega}$ and
$\frac{d\sigma_\perp}{d\Omega}$ denote the cross sections for linear
photon polarization with polarization vector 
$ \mbox{\boldmath $\epsilon_\|$} = \bf
 p \times (p' \times p)/|p \times ( p'
\times p)|$ in the scattering plane and with polarization vector
$ \mbox{\boldmath $\epsilon_\perp$} \bf
= p' \times p/|p' \times p|$ 
perpendicular to it, respectively.

Note that due to the factor $\Delta_j = m_{\tilde{\chi}^\pm_j}^2 -
m_{\tilde{\nu}_e}^2$ the sign of the 
asymmetry $a_l$ and the azimuthal dependence  of the cross section is
determined by the chargino-sneutrino mass difference. Especially for
an invisible sneutrino it is always positive.

The best source of high energy photon beams is that of Compton
backscattered intense laser pulses off one of the beams of a 
linear collider in the $e^+e^-$
or $e^-e^-$ mode. For longitudinally polarized electrons with
energy $E_e$ and helicity $\bar{\lambda}_e$ and circularly polarized
laser photons with energy $E_L$ and helicity $\lambda_L$ the photon
energy spectrum is given by \cite{ginzburg,borden,cuypers2,cuypers1}
\begin{equation} \label{py}
 P(y,\bar{\lambda}_e,\lambda_L) 
 	 = \frac{1}{\sigma_c} \frac{2\pi\alpha^2}{x m_e^2}  \bigg[ 
      \frac{1}{1-y} + 1-y - 4r(1-r) 
        -\bar{\lambda}_e\lambda_L rx(2r-1)(2-y) \bigg]
\end{equation}
with $y=E_\gamma/E_e$, where $E_\gamma$ is the energy of the high
energy photons, $x  \equiv  \frac{4 E_e E_L}{m_e^2}$ and $r \equiv
\frac{y}{x(1-y)}$. 
The total Compton cross section
\begin{equation}
  \sigma_c  =  \sigma_c^0 + \bar{\lambda}_e\lambda_L \sigma_c^1 
\end{equation}
with
\begin{equation}
 \sigma_c^0  = 
     \frac{\pi\alpha^2}{x m_e^2}  \bigg[ 
	\left(2-\frac{8}{x}-\frac{16}{x^2} \right) \ln(x+1) 
    +1+\frac{16}{x}-\frac{1}{(x+1)^2} \bigg],
\end{equation}
\begin{equation}	
  \sigma_c^1  =  \frac{\pi\alpha^2}{x m_e^2} \left[
	\left(2+\frac{4}{x} \right) 
  		\ln(x+1) -5+\frac{2}{x+1}-\frac{1}{(x+1)^2}  \right] 
\end{equation}
normalizes the distribution to $\int P(y)\,dy = 1$.
The fraction $y=E_\gamma/E_e$ of the photon and $e^\pm$ beam energies
varies in the range
\begin{equation}
0 \le y \le \frac{x}{x+1} \mbox{\quad with\quad}
x<2(1+\sqrt{2})\approx 4.83 .
\end{equation}
For higher values of $x$ the conversion efficiency would be
drastically reduced by production of $e^+e^-$ pairs in collisions of
high energy photons and laser photons.

Polarizing the electrons and/or the laser photons results in polarized
backscattered  photons with mean helicity
\begin{equation} \label{lambday}
   \lambda(y,\bar{\lambda}_e,\lambda_L) =  
 \frac{\lambda_L (1-2r)(1-y+\frac{1}{1-y}) + \bar{\lambda}_e
	rx[1+(1-y)(1-2r)^2]} 
     {\frac{1}{1-y} + 1-y - 4r(1-r) - 
     \bar{\lambda}_e\lambda_L rx(2r-1)(2-y)}.
\end{equation}
If the laser light has linear polarization $P_{tL}$ then for
unpolarized electrons the high energy
photons are linearly polarized in the same direction
as the laser light. Their degree of polarization is
\begin{equation} 
  P_t(y) = \frac{2 r^2 P_{tL}}
  	{\frac{1}{1-y} + 1-y - 4r(1-r)}
\end{equation}
and the energy spectrum is given by Equ.~(\ref{py}) with
$\bar{\lambda}_e = \lambda_L = 0$.

To obtain the total cross section $\sigma_{ee}$ and the polarization
asymmetry $A_{ee}$ in the laboratory frame we fold the $e\gamma$-cross
section equ.~(\ref{totwq}) with the energy distribution
equ.~(\ref{py}): 
\begin{equation}  
  \sigma_{ee}(s_{ee},\bar{\lambda}_e,\lambda_L) 
	= \int\!dy\,P(y,\bar{\lambda}_e,\lambda_L) 
	     \sigma(s_{e\gamma},y,\bar{\lambda}_e,\lambda_L)
\end{equation}
with
\begin{equation} 
\sigma(s_{e\gamma},y,\bar{\lambda}_e,\lambda_L) 
	 =  \sigma^0(s_{e\gamma}) (1-\lambda_e) 
	 \left\{ 1+\lambda(y,\bar{\lambda}_e,\lambda_L))
			A_c(s_{e\gamma}) \right\} \nonumber
\end{equation}
and
\begin{equation}  s_{e\gamma} = y s_{ee}. \end{equation}
Then the polarization asymmetry of the convoluted cross section
\begin{equation}
 A_{ee}(\bar{\lambda}_e,\lambda_L) = 
	\frac{\sigma_{ee}(s_{ee},\bar{\lambda}_e,\lambda_L) -
	      \sigma_{ee}(s_{ee},-\bar{\lambda}_e,-\lambda_L)}{
	      \sigma_{ee}(s_{ee},\bar{\lambda}_e,\lambda_L) +
	      \sigma_{ee}(s_{ee},-\bar{\lambda}_e,-\lambda_L)}
\end{equation}
becomes
\begin{equation} 
 A_{ee}(\bar{\lambda}_e,\lambda_L) 
   =	\frac{\int\!dy\,P(y,\bar{\lambda}_e,\lambda_L)
	      \lambda(y,\bar{\lambda}_e,\lambda_L) \sigma^0(ys_{ee})
	      A_c(ys_{ee})}
	{\int\!dy\,P(y,\bar{\lambda}_e,\lambda_L) \sigma^0(ys_{ee})}.
\end{equation}

\section{Numerical results and discussion}
\label{sec3}
\subsection{Scenarios}

For the numerical calculations we have chosen three characteristic
scenarios of gaugino-higgsino mixing. Table \ref{scentab} shows the
masses of the charginos and the two lightest neutralinos and the
mixing parameters $V_{ij}$ of the charginos \cite{bartlfraascharg}
relevant for 
their coupling to electrons and sneutrinos. Also shown is the state of 
the lightest neutralino $\tilde{\chi}^0_1$ in the basis of the weak
eigenstates 
$(\tilde{\gamma},\tilde{Z},\tilde{H}^0_a,\tilde{H}^0_b)$. Generally
the masses and couplings of the charginos and neutralinos depend on
the four parameters $M$, $M'$, $\mu$ and $\tan\beta$ 
\cite{bartlfraascharg,bartlfraasneut}. As
usual $M'$ is fixed by $M'/M = 5/3 \tan^2\theta_W$ and since in most
cases the dependence on the parameter $\tan\beta$ is rather weak, we
have chosen $\tan\beta = 2$. The parameters $M$ and $\mu$ have been
chosen such, that in all three cases the mass of the lightest
chargino $\tilde{\chi}^\pm_1$ is about 88~GeV. The main difference
between 
the three scenarios is the state of the
lightest neutralino. In scenario A the LSP is almost a photino, in
scenario C it is nearly a pure higgsino and in scenario B it is a
photino-zino-higgsino mixture.

Assuming renormalization group equations, the sfermion masses are
related to the SUSY parameters by \cite{ee500,hallpolchinski}
\begin{equation}
 m^2_{\tilde{f}_{L/R}}  =  m_f^2 + m_0^2 + C(\tilde{f}) M^2 
	\pm m_Z^2 \cos2\beta
                 (T_{3f} - e_f \sin^2\theta_W),
\end{equation}
where $m_0$ is the common scalar mass at the unification point,
$T_{3f}$ and $e_f$ are the third component of the weak isospin and the
charge of the corresponding left- or right-handed fermion,
respectively, and
$C(\tilde{l}_R) \approx 0.23$, $C(\tilde{l}_L) = C(\tilde{\nu})
\approx 
0.79$, $C(\tilde{q}_L) \approx 10.8$, $C(\tilde{q}_R) \approx 10.1$.

Then for any value of $m_0$ there exist certain regions in the
parameter space where the sneutrino is lighter than the lightest
chargino $\tilde{\chi}^\pm_1$ and the second lightest neutralino
$\tilde{\chi}^0_2$. In 
these regions only the invisible
decay $\tilde{\nu}_e \to \nu_e + \tilde{\chi}^0_1$ is allowed leading
to one-sided 
events with the detailed signatures determined by the branching
ratios for the chargino decay channels.
 
Fig.~\ref{invsn} shows these regions of an invisible sneutrino for
the two values $m_0 =$ 50 GeV and $m_0 = 100$ GeV of the scalar mass
and $\tan\beta = 2$. For small values of
$m_0$ and especially for $\mu < 0$ the sneutrino will decay invisibly
in a considerable domain of the $M$-$\mu$ plane. With increasing $m_0$ 
these regions are shifted to unphysical large values of $M$ and
$\mu$. Also shown are the parameter space excluded by the chargino
mass 
bound 65~GeV recently published by LEP1.5 \cite{rolandi} 
and the position of 
the scenarios A, B and C. For $m_0 \lesssim
76.8$ GeV the sneutrino is invisible in scenario A.

\subsection{Results for monochromatic photons}
In the following cross sections are given for unpolarized
electrons. Due to the factor $(1-\lambda_e)$ they will be enhanced by 
a factor 2 for left-polarized electron beams ($\lambda_e = -1$).

Fig.~\ref{twqA50} displays the total cross sections
for unpolarized electrons and for different
photon polarizations for the production of the
light chargino $\tilde{\chi}^-_1$ in scenario A for $m_0 = 50$
GeV. $\sigma_+$ and $\sigma_-$ denote the cross sections for photon
helicity 
$\lambda_\gamma = \pm 1$, respectively. For comparison we have added
the 
$\theta$-integrated cross sections for linearly polarized photons,
multiplied by $2\pi$ ($\sigma_{\|/\perp} \equiv 2\pi\,
d\sigma_{\|/\perp}/d\phi$). For
CMS-energies $\sqrt{s} \lesssim  400$ GeV the
cross section is the highest for 
$\lambda_\gamma = -1$  whereas for $\sqrt{s}
\gtrsim 400$ GeV it is the 
highest for $\lambda_\gamma = +1$. Since in scenario A the cross
section for the heavy chargino is smaller than \mbox{$0.002$ pb} we
omit the 
graph for $\tilde{\chi}^-_2$-production.

In scenario B the cross sections for $\tilde{\chi}^-_1$-production
(fig.~\ref{twqB50}) is of the same order of magnitude as in scenario 
A\@. Again for $\sqrt{s} \lesssim 450$ GeV it is
the largest for 
$\lambda_\gamma = -1$ whereas for higher energies $\sqrt{s}
\gtrsim 450$ GeV 
$\lambda_\gamma = +1$ leads to larger values. Here due to the larger
wino component of the 
heavy chargino the cross section for $\tilde{\chi}^-_2$ 
is two orders of magnitude higher than in scenario A.

In scenario C (fig.~\ref{twqC50}) finally the
large wino component of the heavy chargino compensates
the mass difference between the heavy and the light one so that apart 
from the different thresholds both cross sections for $e^-\gamma \to
\tilde{\nu}_e\tilde{\chi}^-_1$ and $e^-\gamma \to
\tilde{\nu}_e\tilde{\chi}^-_2$ are of the same 
order.
For high energies the cross section for the heavy chargino even
dominates. 

Varying the scalar mass $m_0$ from $m_0 = 50$ GeV to $m_0 = 200$ GeV
increases the sneutrino mass without changing the sneutrino couplings. 
Apart from the higher threshold the cross sections are from the same
order of magnitude and their energy dependence is very similar for
both values of $m_0$. Even in scenario A with very different sneutrino
mass for $m_0 = 50$~GeV ($m_{\tilde{\nu}_e} = 65.1$~GeV) and $m_0 =
200$~GeV ($m_{\tilde{\nu}_e} 
= 204.3$~GeV) the maximum cross section for
$\tilde{\chi}^-_1$-production with 
unpolarized photons is reduced by only 50~\% and that for circularly
polarized photons with $\lambda_\gamma = -1$ by only 35~\%.
For scenarios B and C and for the heavy chargino the reduction of the
maximum cross section is always less than 50~\%. 
Therefore we renounce the graphs for the case $m_0 = 200$~GeV.

Due to the factor $\Delta_j = m_{\tilde{\chi}^\pm_j}^2 -
m_{\tilde{\nu}_e}^2$ in equ.~(\ref{al}) 
in scenario A for $e^-\gamma \to
\tilde{\chi}^-_1\tilde{\nu}_e$ and in 
scenarios B and C for $e^-\gamma \to \tilde{\chi}^-_2\tilde{\nu}_e$
the asymmetry for 
linearly polarized photons changes sign when we turn from $m_0 =
50$~GeV to $m_0 = 200$~GeV. Quite generally, however, the nature of
the chargino is by far more crucial than the mass of the sneutrino.

To illustrate the influence of wino-higgsino mixing we give in
fig.~\ref{cp1} the contour lines of the cross sections for 
$e^-\gamma \to \tilde{\chi}^-_1\tilde{\nu}_e$ and $e^-\gamma \to
\tilde{\chi}^-_2\tilde{\nu}_e$ for unpolarized photons and both values
of $m_0$.
We have marked in fig.~\ref{cp1}(a) ($m_0 = 50$~GeV) 
the regions of an
invisible sneutrino while for $m_0 = 200$~GeV they are shifted to
large 
values of $M$ and $\mu$ ($M \gtrsim 410$~GeV,
$\mu \lesssim  -500$~GeV for $\mu < 0$ and $M
\gtrsim 420$~GeV, $\mu
\gtrsim 740$~GeV for $\mu > 0$).

The circular polarization asymmetry $A_c$ of the total cross
sections is depicted in fig.~\ref{AcA50} for
scenario A with $m_0=50$~GeV.  
Apart from a 
somewhat different threshold behavior the magnitude as well as the
energy dependence is very
similar for $m_0 = 200$ GeV. In both cases for
the light chargino it is varying between $A_c = -1$ and $A_c = +0.4$
between threshold and $\sqrt{s} = 1000$~GeV. For the
heavy chargino the circular polarization asymmetry 
$A_c$, $< 0$ in this energy region, changes sign
at $\sqrt{s} \approx 1000$~GeV ($A_c \to 1$ for $\sqrt{s}\to\infty$).

Contrary to the circular polarization asymmetry the linear
polarization asymmetry $a_l$ for scenario A is very different  
for $m_0 = 50$~GeV and $m_0 = 200$~GeV (fig.~\ref{AlA50}).  
Due to the factor
$\Delta_j$ in
equ.~(\ref{al}) it is positive  
for the heavy chargino for both values of $m_0$
whereas for the light chargino $a_l > 0$ if the
sneutrino decays invisible ($m_0 = 50$~GeV) and $a_l < 0$ for $m_0 =
200$~GeV when it is heavier than the light chargino ($a_l \to 0$ for
$\sqrt{s}\to\infty$). 

We omit the graphs for the polarization asymmetries for scenarios B
and C. Those for the circular polarization asymmetry are practically
identical with those for scenario A apart from different threshold
energies. For the light chargino the behavior of the linear
polarization asymmetry $a_l$ differs inappreciable from that in
scenario A for the case of a visible sneutrino ($m_0=200$~GeV). For
the heavy chargino the linear polarization asymmetry for $m_0=50$~GeV
is 
less than 15~\% for scenario B, less than 10~\% for scenario C and for
$m_0=200$~GeV it is less than 5~\% in scenario C. 
Because of $m_{\tilde{\nu}_e} \approx m_{\tilde{\chi}^\pm_2}$ 
in scenario B with $m_0=200$~GeV the linear polarization asymmetry
$a_l$ is very small ($<1~\%$).

\subsection{Results for an \boldmath $ee$ \unboldmath collider}

Before entering the discussion of polarization effects we show in
fig.~\ref{eecp} the contour lines of the convoluted
cross sections for unpolarized beams with CMS-energy $\sqrt{s_{ee}} = 
500$~GeV. Since the reaction is particularly suitable for 
the detection of invisible
sneutrinos we have chosen $m_0 = 50$~GeV for the scalar mass. Then in
a large region of the parameter space (see fig.~\ref{cp1}(a)) only the
invisible decay of the sneutrino into the LSP is kinematically
allowed. 
As can be inferred from fig.~\ref{cp1} increasing
$m_0$ will shift the threshold to higher
energies and reduce the convoluted cross sections
between 30~\% and 50~\% for $m_0 = 200$~GeV. 

Corresponding to the
energy $\frac{x}{x+1} E_e \sim 0.83 E_e$ of the hardest photons the
kinematical accessible $M$-$\mu$ region is somewhat smaller as in
fig.~\ref{cp1}. This results in a reduction of the
cross sections for $ee \to \tilde{\nu}_e\tilde{\chi}^-_j$ and
$\sqrt{s_{ee}} = 
500$~GeV compared to those for $e\gamma \to
\tilde{\nu}_e\tilde{\chi}^-_j$ and 
$\sqrt{s_{e\gamma}} = 500$~GeV for scenarios close by the kinematical
boundary in the $M$-$\mu$ plane. Thus the cross section for production
of heavy charginos in scenario C is reduced by more than one order of
magnitude while those for the light chargino are of the
same order in all three scenarios.

Fig.~\ref{ntwqA50} shows the energy dependence of the convoluted
cross section for scenario A with $m_0=50$~GeV. For comparison we have
included also that for $e^-\gamma \to \tilde{\chi}^-_1 \tilde{\nu}_e$
for unpolarized 
photons. Notice that due to the energy distribution of the
backscattered photons for energies $\sqrt{s_{ee}}$ high enough the
convoluted cross section is larger than that for $e^-\gamma \to
\tilde{\chi}^-_1 \tilde{\nu}_e$ for the same $e\gamma$-energy
$\sqrt{s_{e\gamma}}=\sqrt{s_{ee}}$. The background processes displayed
in fig.~\ref{ntwqA50} will be discussed in the next section.

Both the energy spectrum  and the mean helicity
(equs.~(\ref{py}) and (\ref{lambday})) of the high energy photons 
sensibly depend on the helicity $\lambda_L$ of the laser photons and
the helicity $\bar{\lambda}_e$ of the converted electrons
\cite{borden}. 
We therefore discuss in the following the asymmetries
$A_{ee}(\bar{\lambda}_e,\lambda_L)$ of the convoluted cross sections
for the production of the light chargino in scenario A with $m_0 =
50$~GeV for several polarization configurations $(\bar{\lambda}_e ,
\lambda_L)$.

Since due to missing energy it is impossible to reconstruct from the
momenta of the decay products of the supersymmetric particles the
production plane we renounce the discussion of convoluted cross
sections for linearly polarized photons. For a significant discussion
one has to include the decays of the chargino and to study, e.g., the
linear polarization asymmetry of the observable leptons. This is,
however, beyond the scope of our study.

Generally one obtains large negative polarization asym\-me\-tries for
lower 
energies near the threshold and smaller positive asymmetries for
higher energies. The energy region where the asymmetry changes sign
and is therefore small depends on the polarization configuration.
If only the electron beam is polarized (fig.~\ref{nasym10}(a)), the
asymmetry 
attains $-65 \%$ for $\sqrt{s_{ee}} = 200$~GeV and is of the order of
$+15 \%$ for the highest energy $\sqrt{s_{ee}} = 1000$~GeV. In the
energy region around 500~GeV the asymmetry is small. Apart from
shifting the energy region where the asymmetry changes the sign the
situation is very similar if additionally the laser photons are
polarized (figs.~\ref{nasym10}(b), (c)). In
fig.~\ref{nasym10}(b) the steep
descent of the asymmetry just beyond threshold originates from
the special energy dependence of the mean helicity of the
backscattered photons. For $\bar{\lambda}_e = \lambda_L = +1$ is
$\lambda(y,\bar{\lambda}_e,\lambda_L) = +1$ over almost the whole
spectrum with a steep descent to $\lambda(y,\bar{\lambda}_e,\lambda_L)
= -1$ close to the high energy end \cite{borden}.
With exception of the region very close to threshold the asymmetry
for unpolarized electrons and polarized photons (fig.~\ref{nasym10}(d))
is 
substantially 
smaller with values between $-20 \%$ for $\sqrt{s_{ee}} = 200$~GeV and
$+20 \%$ for $\sqrt{s_{ee}} = 340$~GeV.

The discussion of the polarization  asymmetries  for the background
processes displayed in fig.~\ref{nasym10} shall be
postponed 
to the following section.

\section{Signatures and Background}

For the feasibility of the detection of SUSY signals the decay
patterns of the produced particles are as crucial as their production
cross sections. Our investigations should therefore be completed by an
analysis of the decay channels and a study of the standard model
background. For the present we shall however restrict ourselves to
some remarks emphasizing the usefulness of polarized beams.
For scenario A with $m_0 = 200$~GeV and $m_{\tilde{\nu}_e} =
204.3$~GeV 
as well as for scenarios B and C with heavy sneutrinos the dominating
signature with branching ratio between 35~\% and 40~\% are four jets
accompanied by one electron (or one positron) via
\begin{center}
\setlength{\unitlength}{2em}
\begin{picture}(8.5,2.5)
\put(.1,2){$e^- + \gamma \longrightarrow \tilde{\nu}_e +
\tilde{\chi}^-_1$} 
\put(3.9,1.5){\line(0,1){0.25}}
\put(3.86,1.39){$\longrightarrow \bar{u} + d + \tilde{\chi}^0_1$}
\put(2.8,0.9){\line(0,1){0.85}}
\put(2.76,0.79){$\longrightarrow e^- + \tilde{\chi}^+_1$}
\put(5.05,0.3){\line(0,1){0.25}}
\put(5.01,0.19){$\longrightarrow u + \bar{d} + \tilde{\chi}^0_1$ . }
\end{picture}
\end{center}
The decay rates and the branching ratios of the supersymmetric
particles have been calculated with the formulae given in
refs.~\cite{diswoehrmann,bartlfraaszerfaelle}.
 
The standard model background for this signature resulting from
higher-order processes is expected to be rather small. Since, however, 
a heavy sneutrino decaying visibly should also be observable at an
$e^+e^-$ collider we shall not further consider this case. 
We focus on the particular interesting case of associated
production of a light invisible sneutrino and the light chargino
realized in scenario A with $m_0=50$~GeV. In this case the chargino
almost exclusively decays into the invisible $\tilde{\nu}_e$ and one
electron 
so that for $\sqrt{s_{ee}} = 500$~ GeV the cross section for
$e^-\gamma \to e^- + \mbox{$\not\!\!E$}$ attains values between 1~pb
and 2~pb. 
On account of the small cross section contributions from the
production and decay of the heavy chargino they are negligible.
For this case the $e\gamma$ option
seems to be superior to the radiative process $e^+e^- \to \tilde{\nu} 
\bar{\tilde{\nu}} \gamma$ with cross sections at least two orders of
magnitude smaller \cite{franke}.

The most important background to the signal arises from the two
processes $e^-\gamma \to W^-\nu_e$ and $e^-\gamma \to e^-Z$. 
The cross sections for this processes shown in fig.~\ref{ntwqA50}
are discussed in detail in
\cite{dittmaier,baillargeon,belanger,renard} and yield values of 30~pb
and 8~pb for $\sqrt{s_{ee}} = 500$~GeV, respectively.
This seems very high but the background is reduced to
$\sigma(e \gamma \to (W \nu_e , e Z) \to 1 e^-+ \mbox{$\not\!\!E$} )
\approx$ \mbox{9 pb} for the $(1 e^-+ \mbox{$\not\!\!E$})$-signature.  
Furthermore one expects, that similar as for the process $e\gamma \to
\tilde{e}\tilde{\chi}^0_j$ the supersymmetric signal can be enhanced
relative 
to the SM background by cuts on the rapidity and transverse momentum
of the final state electron \cite{cuypers2}. 

Although the contribution from $e^-\gamma \to e^-Z$ is rapidly
decreasing with increasing energy one should note that it is more
important for the convoluted cross sections than for the nonconvoluted
ones, compared for the same CMS energy $\sqrt{s_{ee}}$ and
$\sqrt{s_{e\gamma}}$, respectively; e.g., one finds
$\sigma(e\gamma \to eZ) \sim 3$~pb (0.8~pb) for
$\sqrt{s_{e\gamma}} = 500$~GeV (1000~GeV) whereas due to folding with
the photon spectrum $\sigma(ee \to e\gamma \to eZ) \sim 8$~pb (3~pb)
for 
$\sqrt{s_{ee}} = 500$~GeV (1000~GeV). 

The elimination of the SM background can considerably be facilitated
by a polarized electron beam (fig.~\ref{nasym10}(a)).
In most of the energy region between threshold and 1~TeV the
polarization asymmetries of the background processes are considerably
smaller or even have different sign than that for the supersymmetric
process. Comparison of fig.~\ref{nasym10}(a) with
figs.~\ref{nasym10}(b), (c)
demonstrates that additional polarization of the laser
photons would not substantially improve the situation. 
Thus combined with cuts in rapidity and transverse momentum the
polarization asymmetries for a polarized electron beam will improve
chances for elimination of the SM background.

Beyond associated production of an invisible sneutrino and the light
chargino the supersymmetric process $e^-\gamma \to
\tilde{e}^-\tilde{\chi}^0_1 
\to e^-\tilde{\chi}^0_1\tilde{\chi}^0_1$ will contribute to the single
electron 
signal. This is on the one hand an additional background for the
process under study, on the other hand it will enhance the
supersymmetric signal relative to the SM background. We shall leave
the study of this process to further investigations.

\section{Summary}

We have calculated the cross sections and polarization asymmetries for
the reaction $e^- \gamma \to \tilde{\chi}^-_j \tilde{\nu}_e$. The high
energy photons 
can be generated in the $e\gamma$ collision mode
by Compton backscattering of intense laser pulses off
one of the beams of an $e^+e^-$ or $e^-e^-$ linear collider. 
To obtain the cross 
sections and polarization asymmetries in the laboratory frame they
must be convoluted with the Compton energy spectrum.

We have chosen three characteristic scenarios of gaug\-ino-hig\-gsino
mixing for the numerical calculations. The cross sections for the
production of the light chargino are of the order of 1~pb in a
considerably domain of the parameter space.

For sneutrinos decaying invisibly into a neutrino and the
LSP the lighter chargino decays through a dominant
two-body decay channel nearly exclusively in a sneutrino and an
electron yielding a clear signature. 
Therefore the reaction $e^- \gamma \to \tilde{\chi}^-_1 \tilde{\nu}_e$
is 
suitable to search for such invisible sneutrinos because here the
cross sections are two orders of magnitude higher than for the
radiative production of the sneutrinos \cite{franke}.

The polarization asymmetries of the SM background processes $e\gamma
\to W\nu_e$, $e\gamma \to eZ$ are smaller or have opposite sign for
most of the energy region. So using polarized beams combined with cuts
in rapidity and transverse momentum can considerably enhance the
signal-to-background ratio.

\acknowledgements

We would like to thank F.~Franke for many helpful comments on the
manu\-s\-cript. This work was supported by the Deutsche
Forschungsgemeinschaft under contract no.\ \mbox{FR~1064/2-1}.

\newpage
\begin{figure}
\vspace*{4cm}
\centering
\setlength{\unitlength}{10mm}
\begin{picture}(13.5,9)
\put(-3.5,-10){\includegraphics{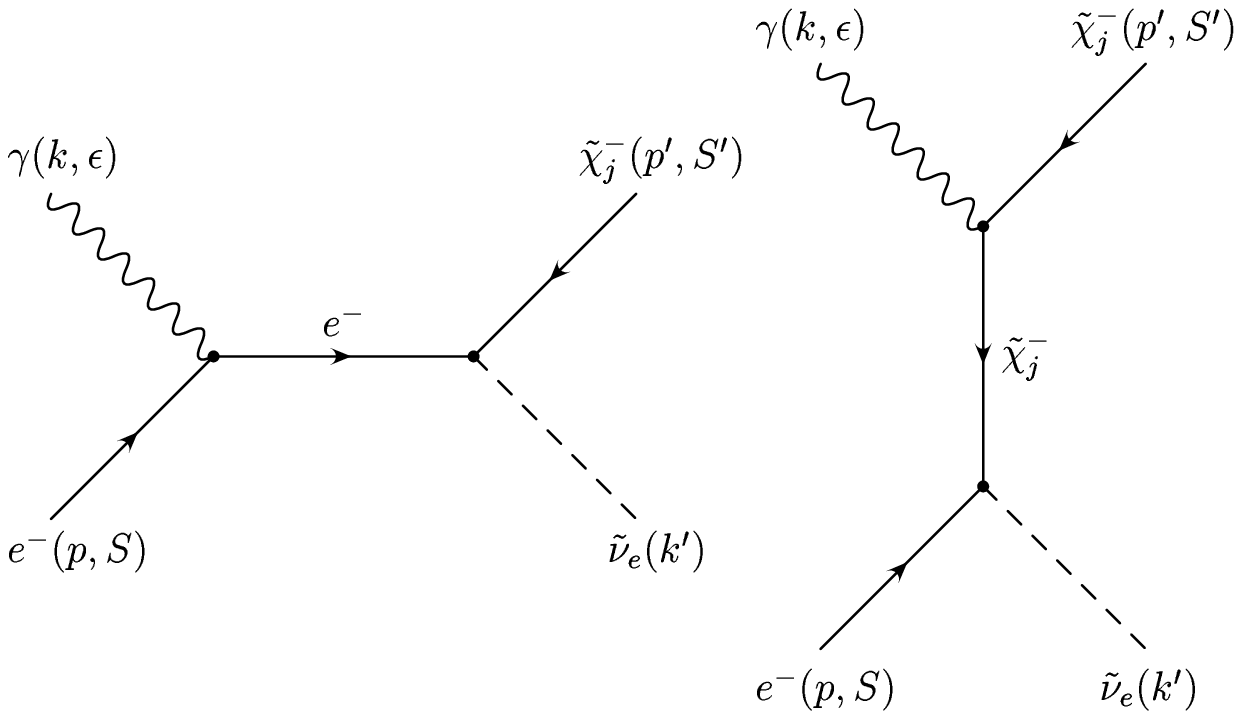}}
\end{picture}
\caption{\label{feyn1} Feynman graphs for $e^- +
      \gamma \rightarrow \tilde{\chi}^-_j + \tilde{\nu}_e $.}
\end{figure}
\clearpage

\vspace*{3cm}
\begin{figure}
\centering
\setlength{\unitlength}{10mm}
\begin{picture}(14,13.5)
\put(-3.3,-6.9){\includegraphics{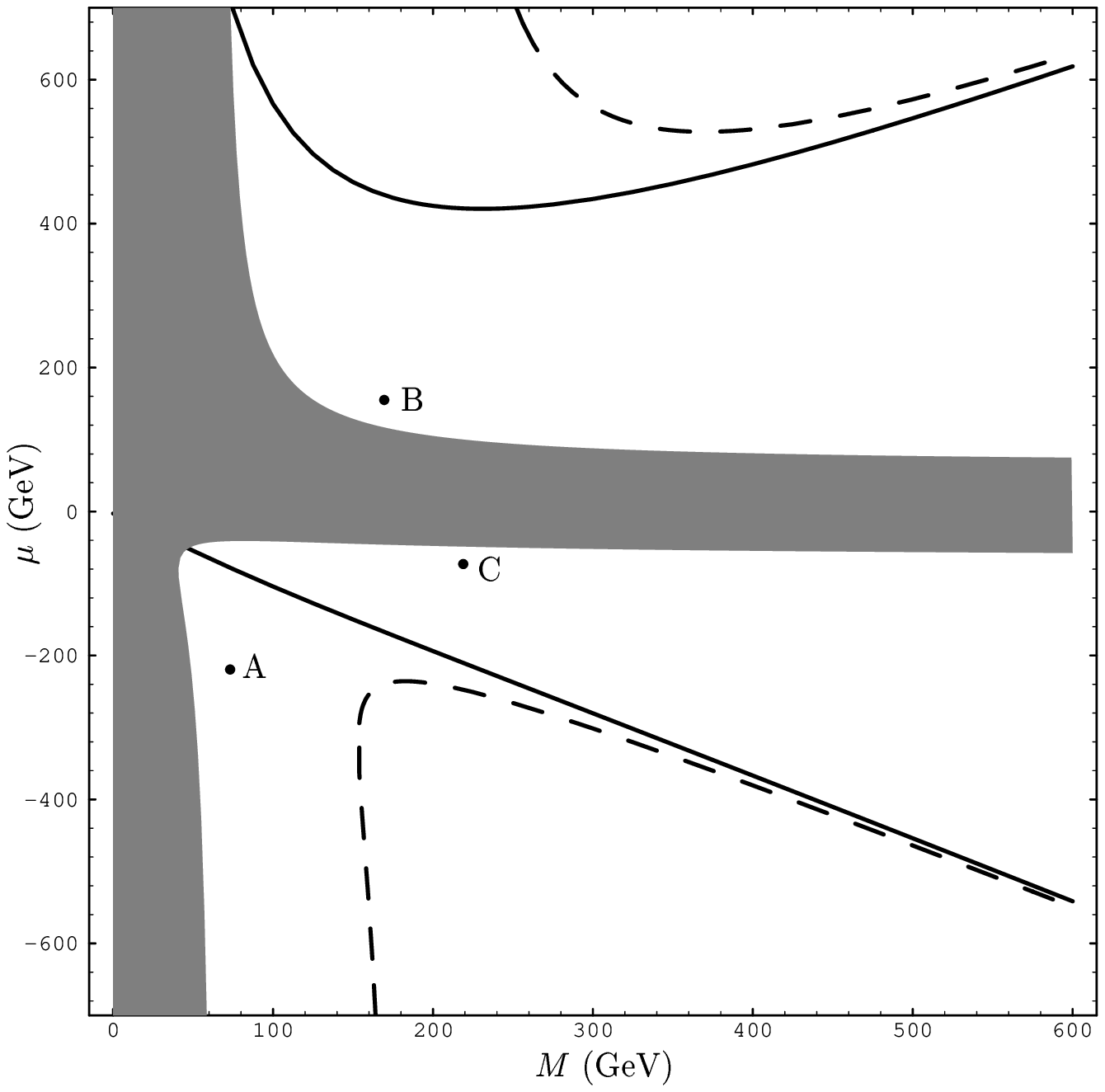}}
\end{picture}
\caption{\label{invsn} Regions of an invisible sneutrino in the
$M$-$\mu$ parameter space for the two
values $m_0 = 50$~GeV (solid line) and $m_0 = 100$~GeV (dashed line)
of the scalar mass and $\tan\beta=2$. 
The sneutrinos are invisible above the upper lines
and below the lower lines, respectively. Also shown are the 
experimentally excluded parameter space (shaded area) and the position
of the three scenarios A, B and C.}
\end{figure}
\clearpage

\begin{figure}
\centering
\setlength{\unitlength}{1cm}
\begin{picture}(14.1,8.8)
\put(-3.4,-9.2){\includegraphics{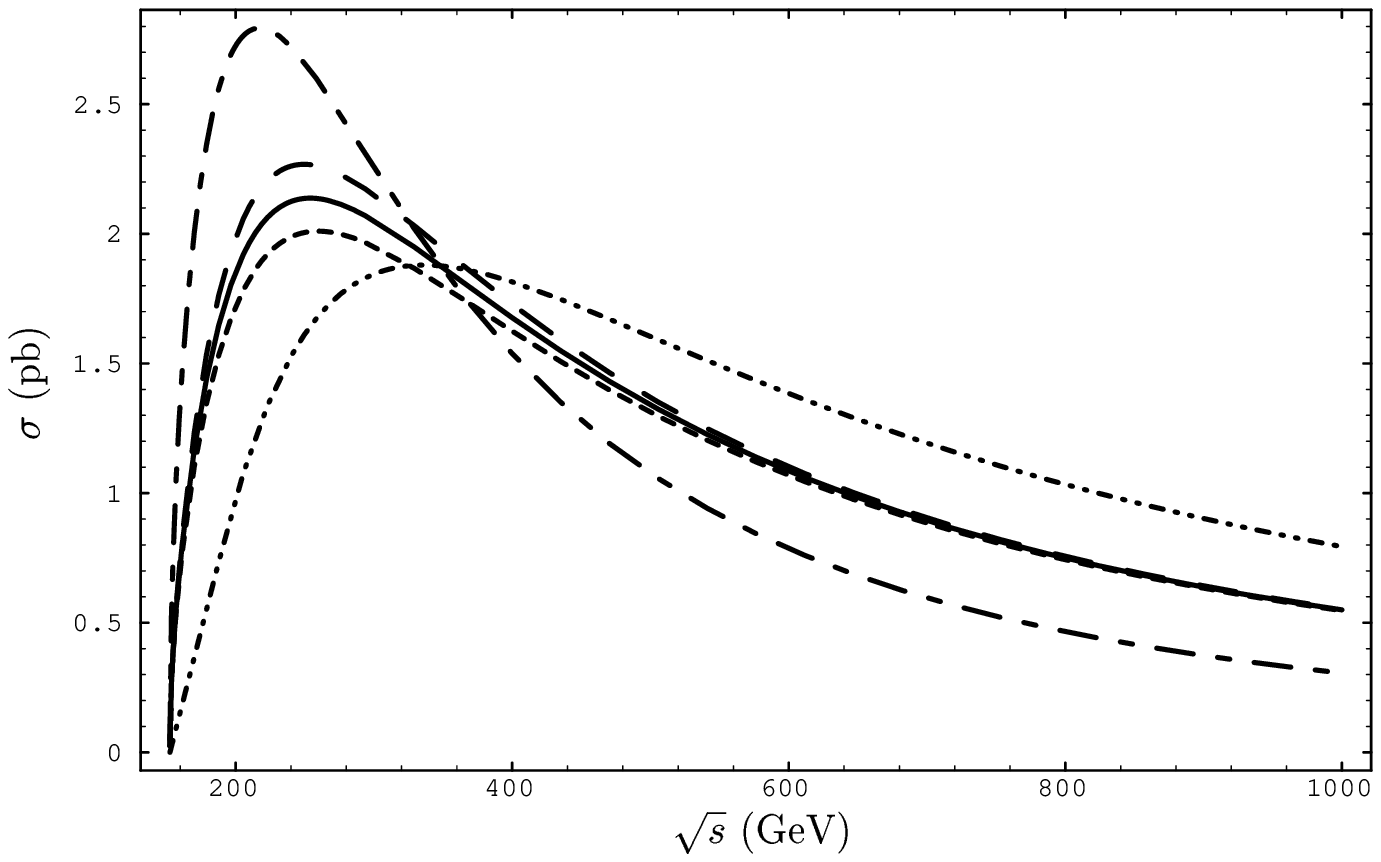}}
\end{picture}
\caption{\label{twqA50} Total cross sections
$\sigma(e^-\gamma\to\tilde{\chi}^-_1\tilde{\nu}_e)$ in scenario A 
for $m_0=50$ GeV: 
\mbox{$\sigma^0$ (---\hspace{-.5mm}---\hspace{-.5mm}---)}, 
\mbox{$\sigma_\|$ (---\quad---)},
\mbox{$\sigma_\perp$ (-- -- -- --)}, 
\mbox{$\sigma_+$ (-- $\cdot$ $\cdot$ --)}, 
\mbox{$\sigma_-$(--- -- ---)}.}
\end{figure}

\begin{figure}
\centering
\setlength{\unitlength}{1cm}
\begin{picture}(14.1,8.8)
\put(-3.4,-9.2){\includegraphics{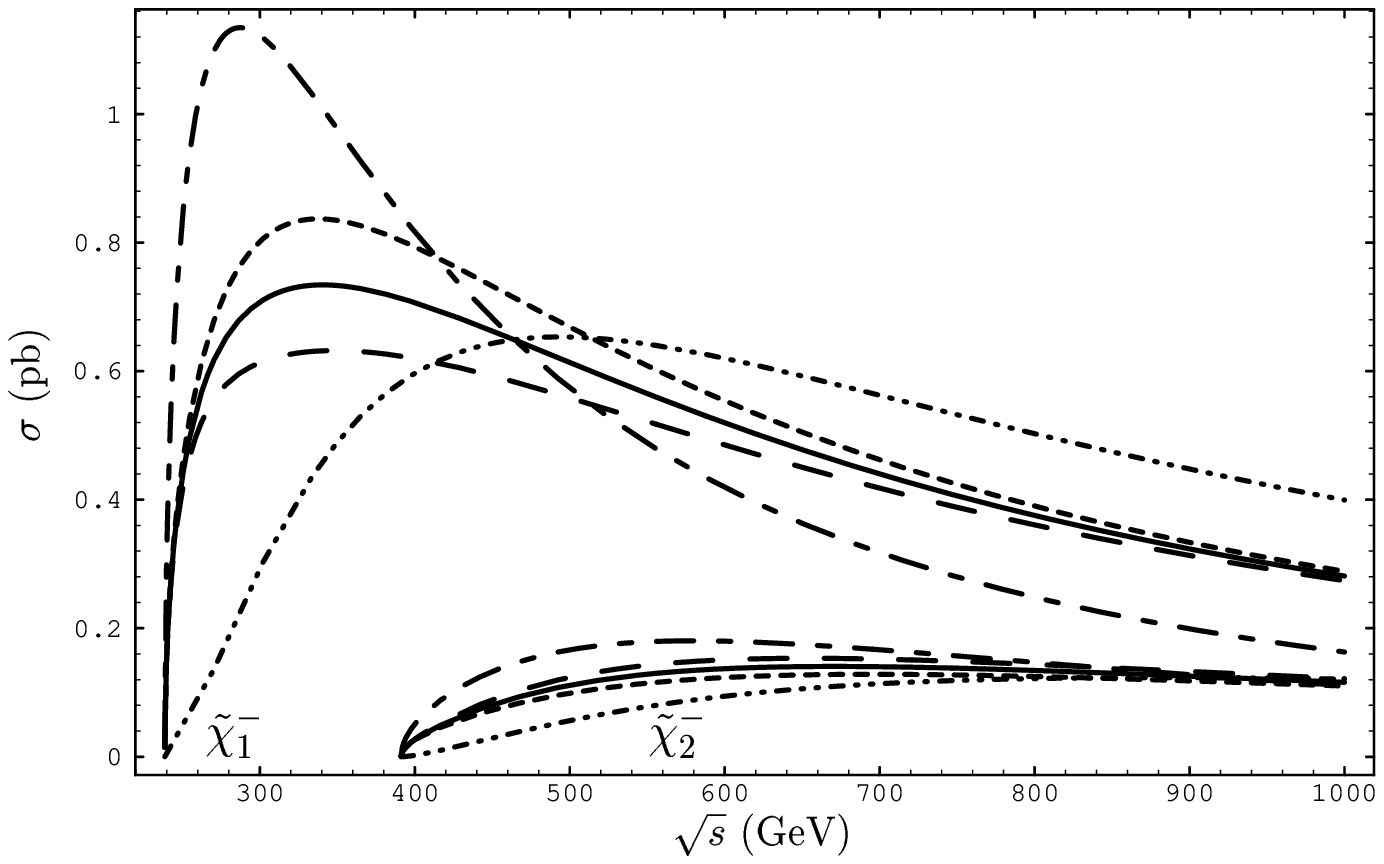}}
\end{picture}
\caption{\label{twqB50} Total cross sections
$\sigma(e^-\gamma\to\tilde{\chi}^-_{1,2}\tilde{\nu}_e)$ in scenario B 
for $m_0=50$ GeV: 
\mbox{$\sigma^0$ (---\hspace{-.5mm}---\hspace{-.5mm}---)}, 
\mbox{$\sigma_\|$ (---\quad---)},
\mbox{$\sigma_\perp$ (-- -- -- --)}, 
\mbox{$\sigma_+$ (-- $\cdot$ $\cdot$ --)}, 
\mbox{$\sigma_-$(--- -- ---)}.}
\end{figure}
\clearpage

\vspace*{5cm}
\begin{figure}
\centering
\setlength{\unitlength}{1cm}
\begin{picture}(14.1,8.8)
\put(-3.4,-9.2){\includegraphics{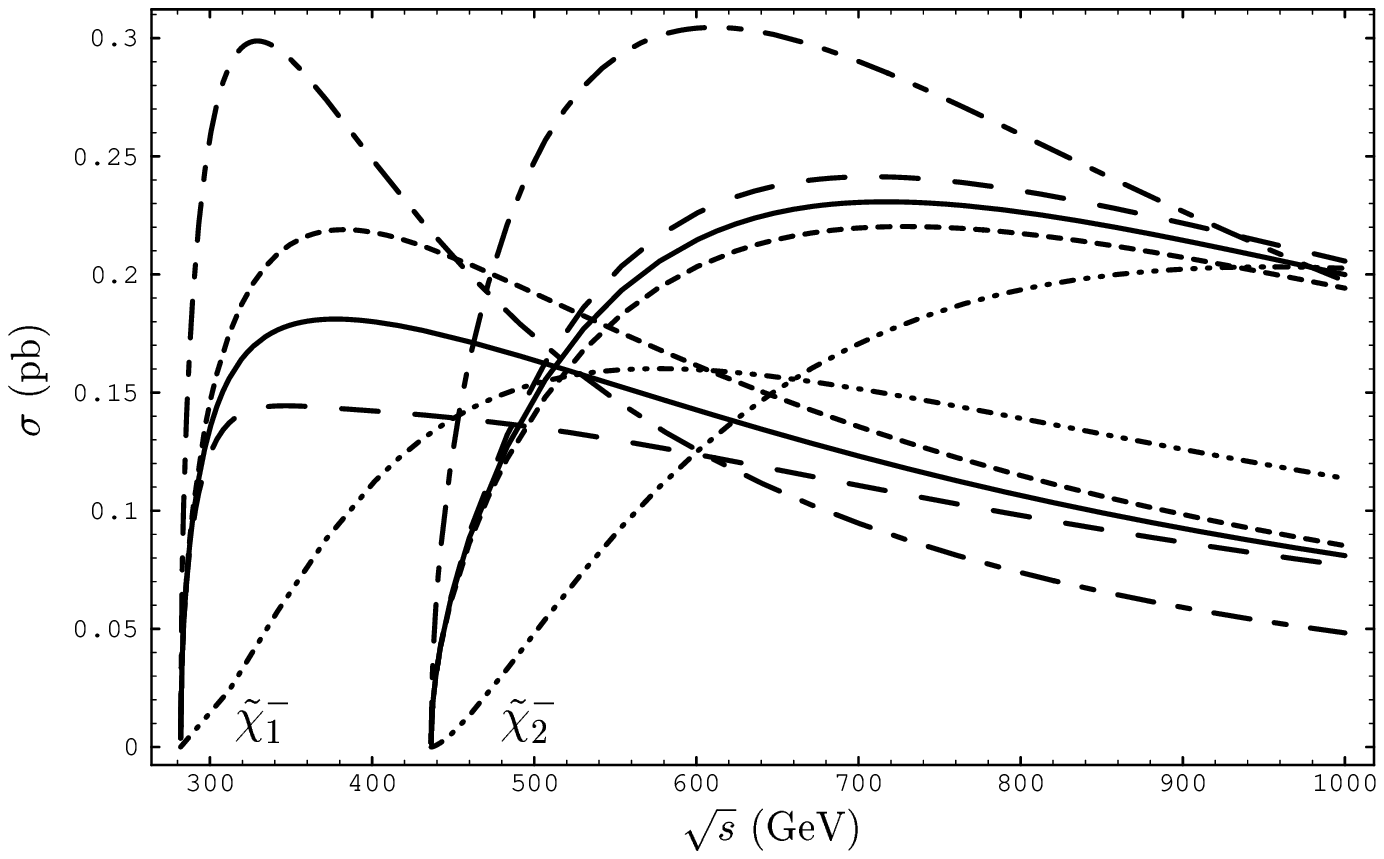}}
\end{picture}
\caption{\label{twqC50} Total cross sections
$\sigma(e^-\gamma\to\tilde{\chi}^-_{1,2}\tilde{\nu}_e)$ in scenario C 
for $m_0=50$ GeV: 
\mbox{$\sigma^0$ (---\hspace{-.5mm}---\hspace{-.5mm}---)}, 
\mbox{$\sigma_\|$ (---\quad---)},
\mbox{$\sigma_\perp$ (-- -- -- --)}, 
\mbox{$\sigma_+$ (-- $\cdot$ $\cdot$ --)}, 
\mbox{$\sigma_-$(--- -- ---)}.}
\end{figure}
\clearpage

\begin{figure}
\centering
\setlength{\unitlength}{1cm}
\begin{picture}(16,17.2)
\put(-3,-8.7){\includegraphics{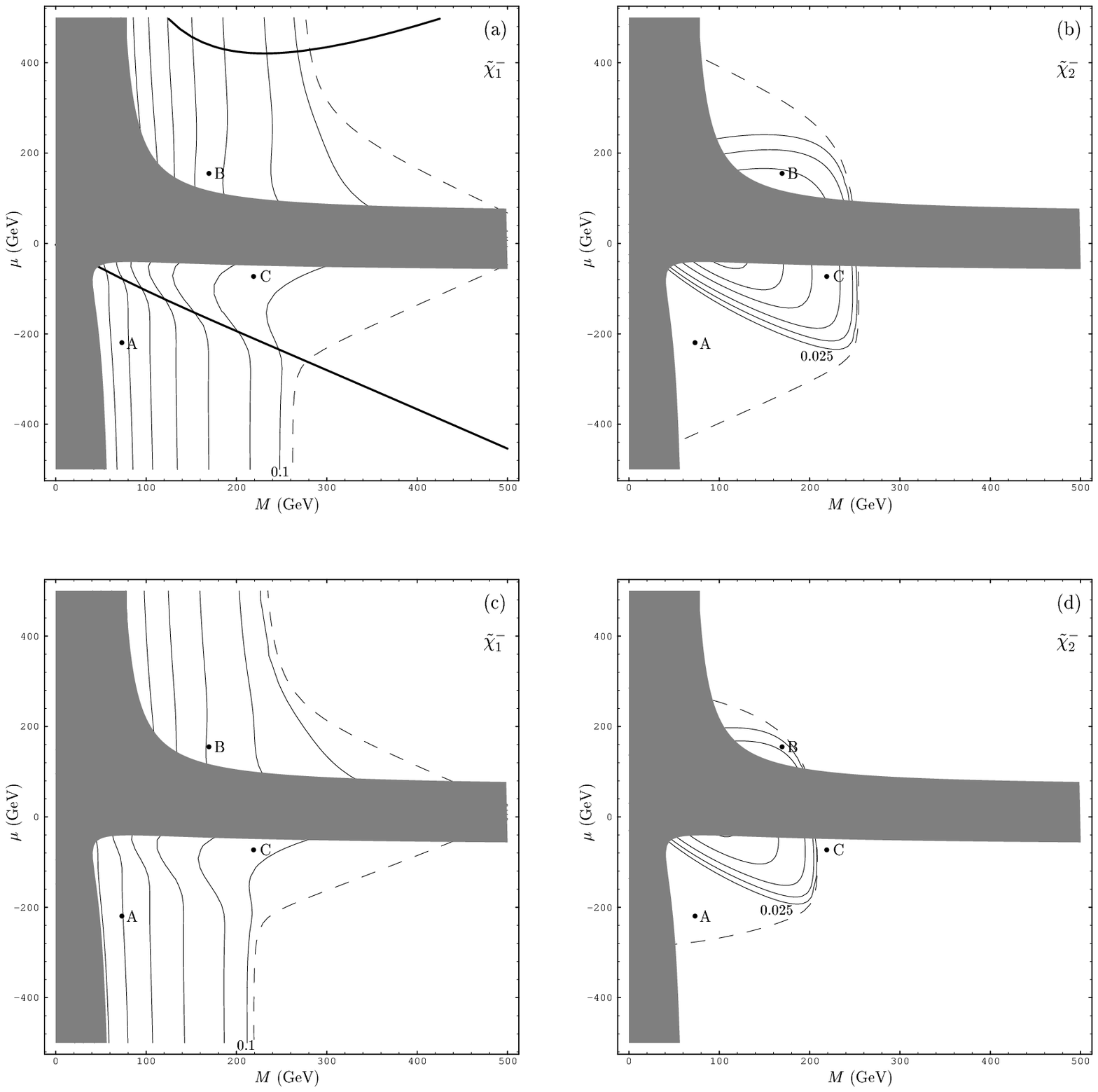}}
\end{picture}
\caption[]{\label{cp1} Contours of the total cross section for
$\sqrt{s} = 500$~GeV and $\tan\beta = 2$. The dashed lines mark the
boundary $m_{\tilde{\chi}^\pm_{1,2}} +m_{\tilde{\nu}_e} = 500$~GeV. 
The experimentally excluded domain is shaded.\\
(a) Contours (0.1~pb, 0.25~pb, 0.5~pb, 0.75~b, 1.0~pb, 
1.25~pb and 1.5~pb) of
$\sigma(e^-\gamma\to\tilde{\chi}^-_1\tilde{\nu}_e)$ for
$m_0=50$~GeV. The thick lines bound the regions of an invisible 
sneutrino similar as in Fig.~\protect\ref{invsn}.\\
(b) Contours (0.025~pb, 0.05~pb, 0.1~pb, 0.2~pb,
0.3~pb and 0.4~pb) of
$\sigma(e^-\gamma\to\tilde{\chi}^-_2\tilde{\nu}_e)$ for
$m_0=50$~GeV.\\
(c) Contours (0.1~pb, 0.25~pb, 0.5~pb, 
0.75~pb, 1.0~pb and 1.25~pb) of
$\sigma(e^-\gamma\to\tilde{\chi}^-_1\tilde{\nu}_e)$ for
$m_0=200$~GeV.\\
(d) Contours (0.025~pb, 0.05~pb, 0.1~pb, 
0.2~pb and 0.3~pb) of
$\sigma(e^-\gamma\to\tilde{\chi}^-_2\tilde{\nu}_e)$ for
$m_0=200$~GeV.} 
\end{figure}
\clearpage

\begin{figure}
\centering
\setlength{\unitlength}{1cm}
\begin{picture}(14.1,8.8)
\put(-3.5,-9.2){\includegraphics{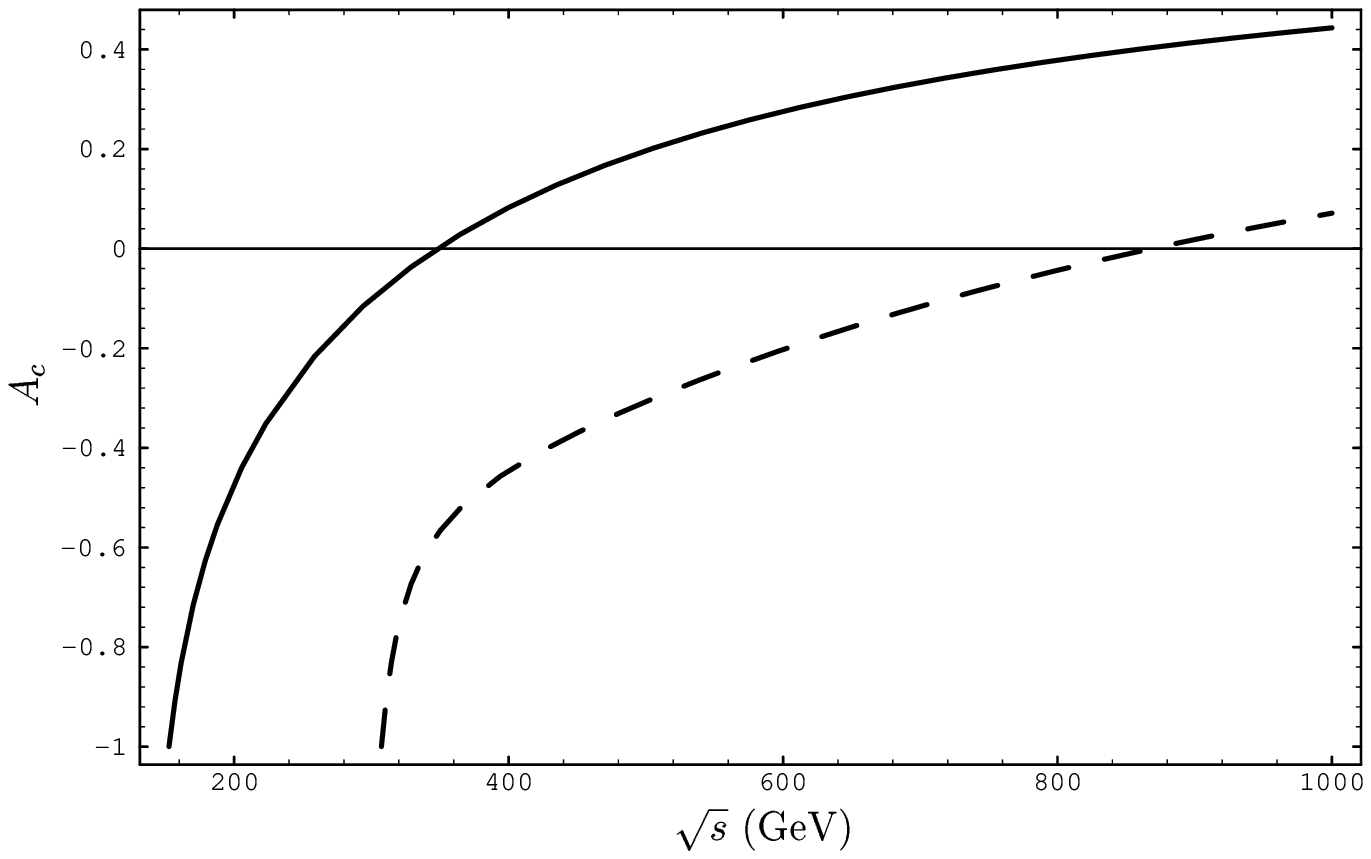}}
\end{picture}
\caption{\label{AcA50} Asymmetry $A_c$ in scenario A with $m_0=50$
GeV for production of $\tilde{\chi}^-_1$
(\mbox{---\hspace{-.5mm}---\hspace{-.5mm}---}) and
$\tilde{\chi}^-_2$ (\mbox{---\quad---}).}
\end{figure}

\begin{figure}
\centering
\setlength{\unitlength}{1cm}
\begin{picture}(14.1,8.8)
\put(-3.5,-9.2){\includegraphics{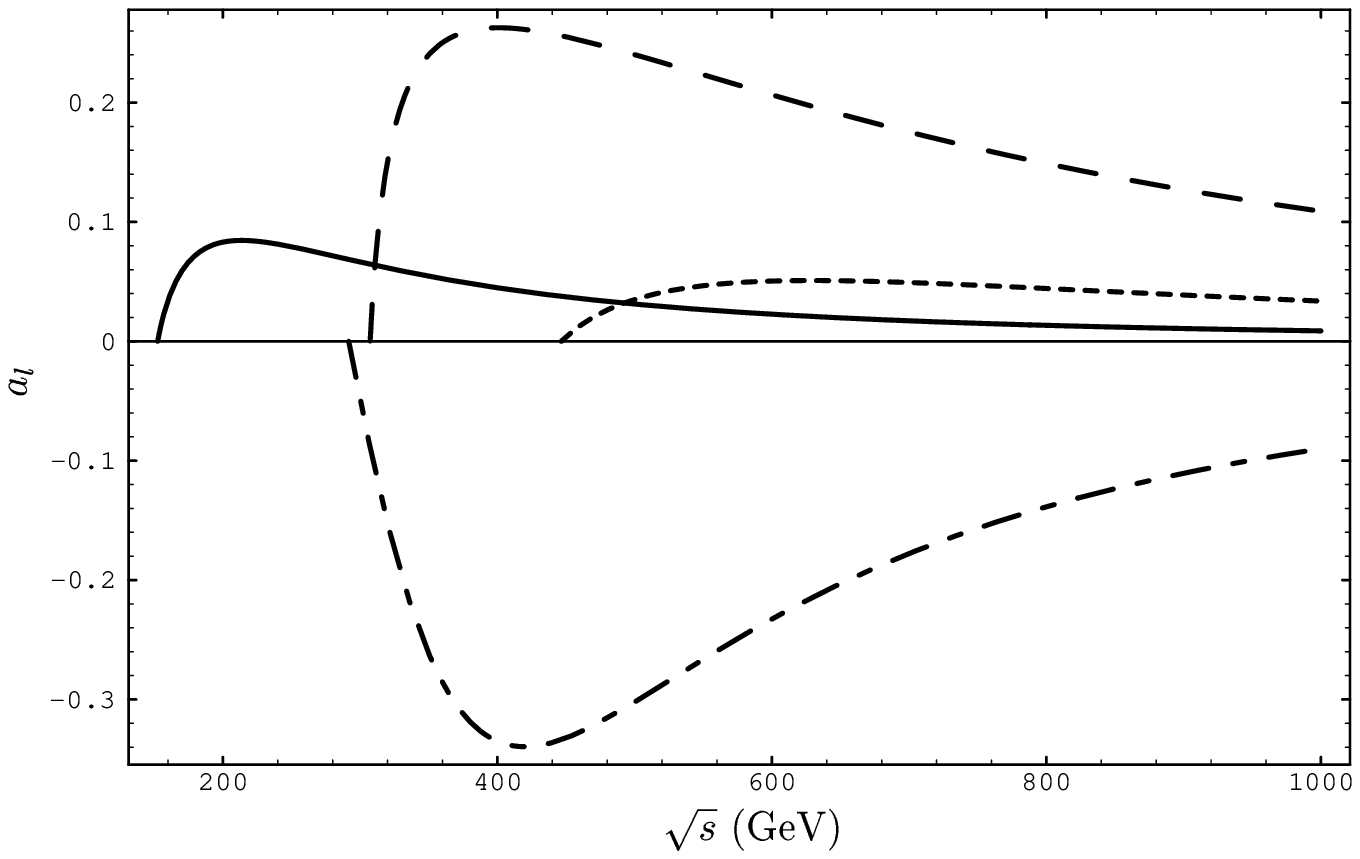}}
\end{picture}
\caption{\label{AlA50} Asymmetry $a_l$ in scenario A with $\cos\theta
= 0.5$ ($\theta=60^\circ$) for production of $\tilde{\chi}^-_1$
(\mbox{---\hspace{-.5mm}---\hspace{-.5mm}---}) and $\tilde{\chi}^-_2$
(\mbox{---\quad---}) with $m_0=50$~GeV and for production of
$\tilde{\chi}^-_1$ (\mbox{--- -- ---}) and $\tilde{\chi}^-_2$
(\mbox{-- -- -- --}) with $m_0=200$~GeV. }
\end{figure}
\clearpage

\begin{figure}
\centering
\setlength{\unitlength}{1cm}
\begin{picture}(8.3,17.2)
\put(-3.1,-8.8){\includegraphics{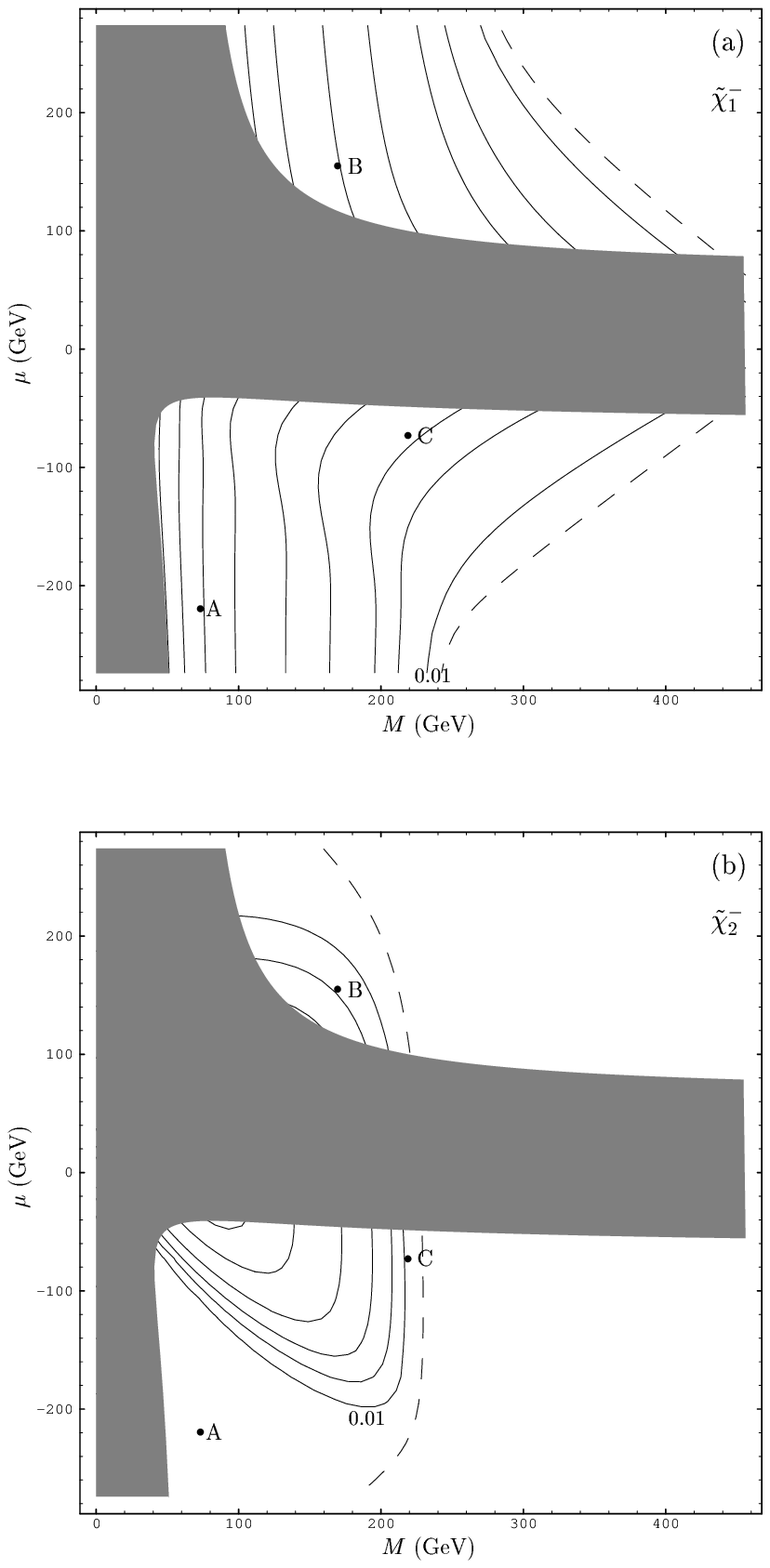}}
\end{picture}
\caption[]{\label{eecp} Contours of the total cross section for
$m_0 = 50$~GeV, $x=4.83$, $\sqrt{s_{ee}}=500$~GeV and
$\tan\beta=2$. The 
dashed lines mark the boundary $m_{\tilde{\chi}^\pm_{1,2}} 
+m_{\tilde{\nu}_e} = 414$~GeV.  
The experimentally excluded domain is shaded.\\
(a) Contours (0.01~pb, 0.05~pb, 0.1~pb, 0.25~pb,
0.5~pb, 1.0~pb, 1.5~pb, 2.0~pb and 2.5~pb) of $\sigma(e
e\to\tilde{\chi}^-_1\tilde{\nu}_e)$.\\
(b) Contours (0.01~pb, 0.025~pb, 0.05~pb, 0.1~pb,
0.2~pb and 0.3~pb) of $\sigma(e e\to\tilde{\chi}^-_2\tilde{\nu}_e)$.}
\end{figure}
\clearpage

\vspace*{5cm}
\begin{figure}
\centering
\setlength{\unitlength}{1cm}
\begin{picture}(14.1,8.8)
\put(-3.3,-9.2){\includegraphics{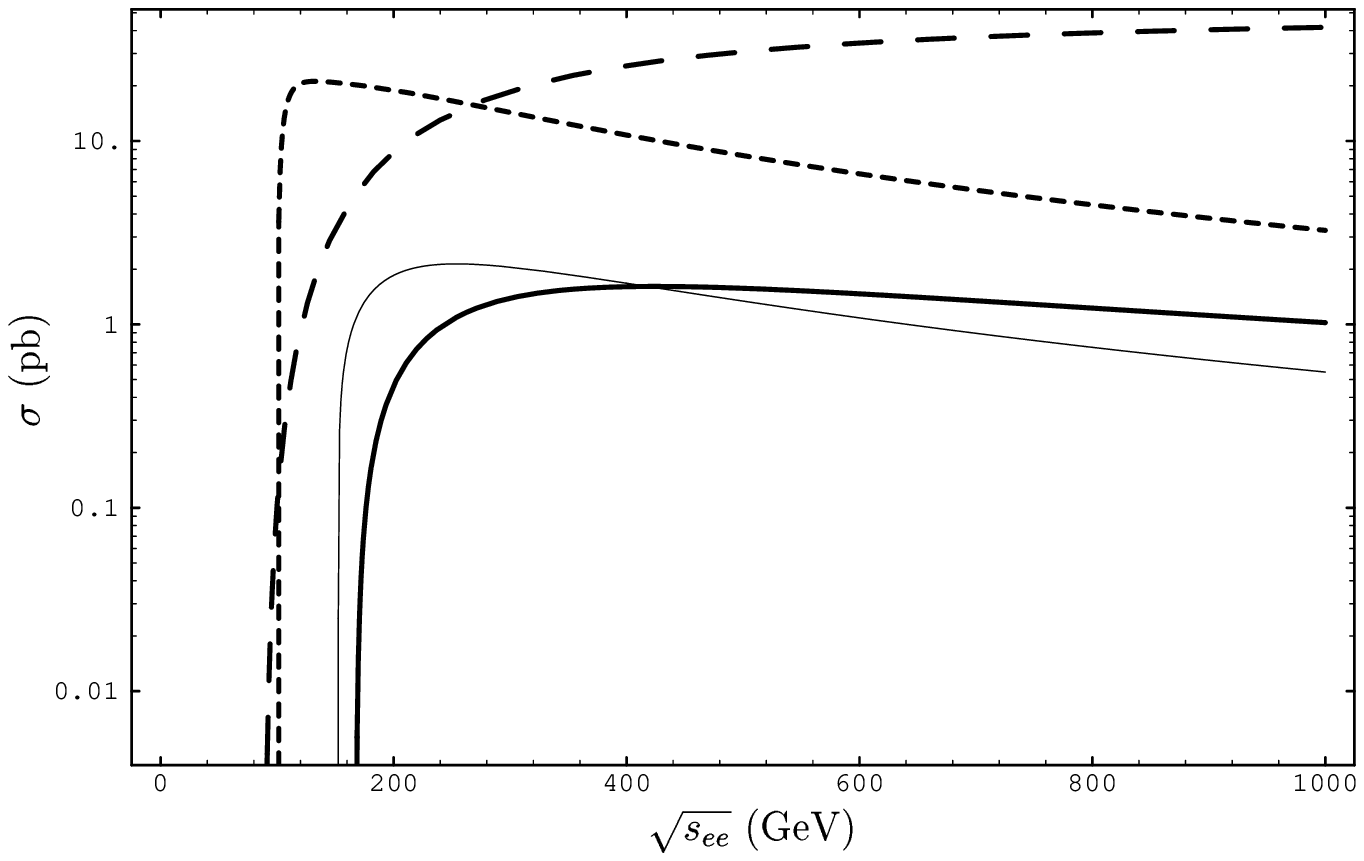}}
\end{picture}
\caption{\label{ntwqA50} Total cross sections $\sigma(s_{ee})$ for
$ee\to\tilde{\chi}^-_1\tilde{\nu}_e$ (\mbox{---\hspace{-.5mm}---}),
$ee \to W\nu_e$  
(\mbox{---\quad---}) and $ee \to eZ$ (\mbox{-- -- -- --})
in scenario A with $m_0=50$~GeV, $x=4.83$ and unpolarized
electron and laser beams.
For comparison we also show $\sigma(s_{e\gamma})$ for
$e\gamma\to\tilde{\chi}^-_1\tilde{\nu}_e$ with unpolarized
beams (thin line).}
\end{figure}
\clearpage

\vspace*{4cm}
\begin{figure}
\centering
\setlength{\unitlength}{1cm}
\begin{picture}(16,11)
\put(-3,-14.9){\includegraphics{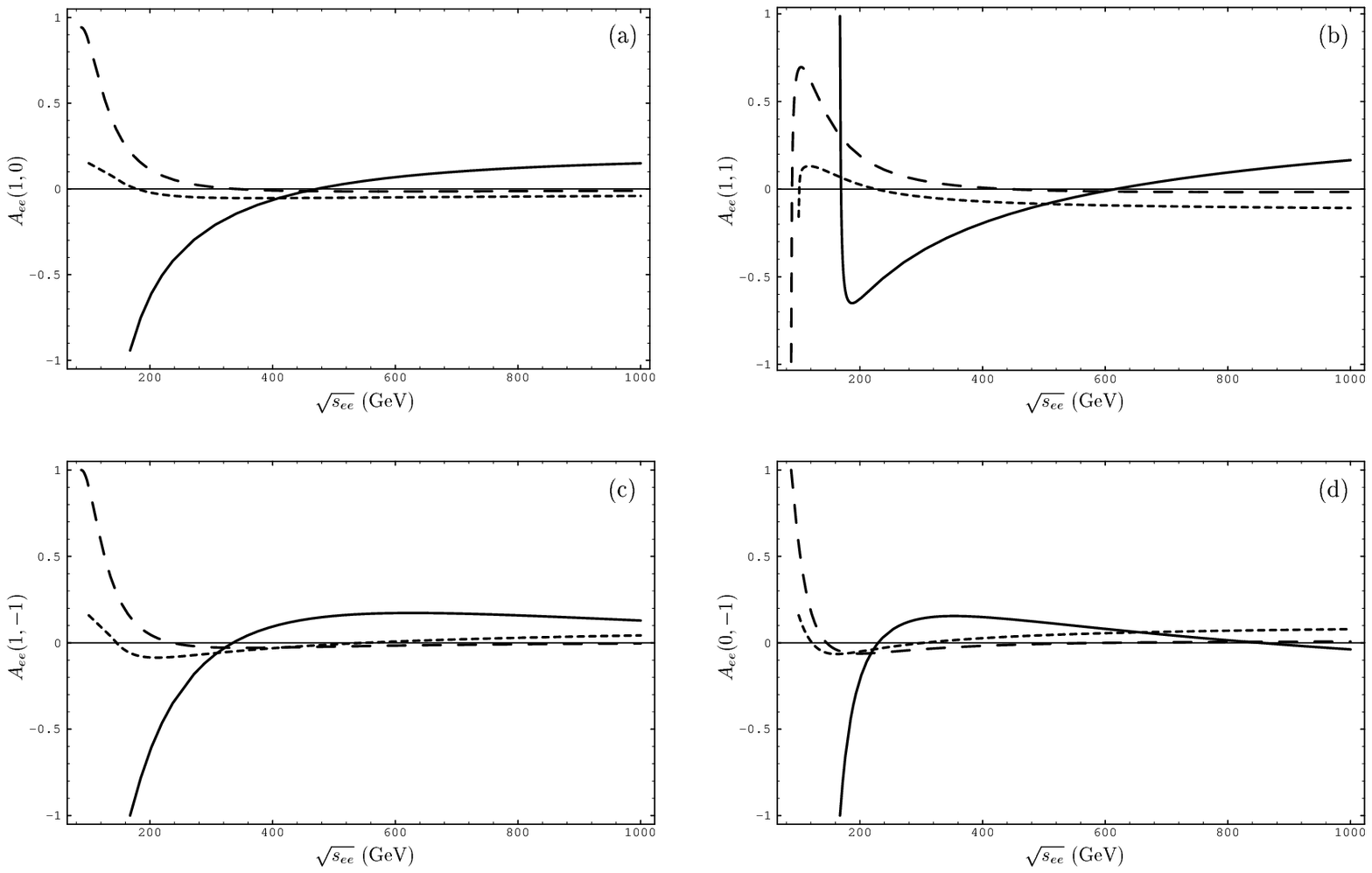}}
\end{picture}
\caption{\label{nasym10} Asymmetries  of
the convoluted cross sections for $ee \to
\tilde{\chi}^-_1\tilde{\nu}_e$ 
(\mbox{---\hspace{-.5mm}---}), $ee \to W\nu_e$ 
(\mbox{---\quad---}) and $ee \to eZ$ (\mbox{-- -- -- --})
in scenario A with $m_0=50$~GeV and $x=4.83$. }
\end{figure}

\mediumtext
\begin{table}
\vspace*{4cm}
\caption{\label{scentab}
Chargino masses $m_{\tilde{\chi}^\pm_j}$ $(j=1,2)$ and
mixing parameters $V_{ij}$  in three different scenarios of the
parameters $M$ and $\mu$. Also shown are the masses of the sneutrinos 
for the two values
of $m_0$ in each scenario, the masses of the lightest and the
second lightest neutralino and the state of the lightest neutralino
$\tilde{\chi}^0_1$ in terms of the weak eigenstates 
$(\tilde{\gamma},\tilde{Z},\tilde{H}^0_a,\tilde{H}^0_b)$.
The $\eta_{\tilde{\chi}^\pm_j}$ and $\eta_{\tilde{\chi}^0_j}$ 
denote the sign of the corresponding eigenvalues of the mass matrices
of charginos and neutralinos, 
respectively.}

\begin{tabular}{lcccccc}
 Scenario & \multicolumn{2}{c}{A} & \multicolumn{2}{c}{B} &
	\multicolumn{2}{c}{C} \\ \hline
 $M$ (GeV) & \multicolumn{2}{c}{$+73.16$} & 
	\multicolumn{2}{c}{$+169.52$} & 
	\multicolumn{2}{c}{$+218.93$} \\
 $\mu$ (GeV) & \multicolumn{2}{c}{$-219.47$} & 
	\multicolumn{2}{c}{$+155.04$} & 
	\multicolumn{2}{c}{$-72.96$} \\ 
 $m_0$ (GeV) & 50 & 200 & 50 & 200 & 50 & 200 \\         \hline 
 $m_{\tilde{\nu}_e}$ (GeV) & 65.1 & 204.3 & 150.7 & 245.4 & 194.6 & 
   274.5 \\ \hline
 $m_{\tilde{\chi}^\pm_1}$, $\eta_{\tilde{\chi}^\pm_1}$ & 
	\multicolumn{2}{c}{87.4 GeV, $+1$} & 
	\multicolumn{2}{c}{88.0 GeV, $-1$} & 
	\multicolumn{2}{c}{87.3 GeV, $+1$} \\
 $m_{\tilde{\chi}^\pm_2}$, $\eta_{\tilde{\chi}^\pm_2}$ & 
	\multicolumn{2}{c}{242.3 GeV, $+1$} & 
	\multicolumn{2}{c}{240.5 GeV, $+1$} & 
	\multicolumn{2}{c}{241.8 GeV, $+1$} \\
 $V_{11} = V_{22}$ & \multicolumn{2}{c}{$+0.9974$} & 
	\multicolumn{2}{c}{$+0.7279$} & 
	\multicolumn{2}{c}{$+0.3956$} \\
 $V_{21} = - V_{12}$ & \multicolumn{2}{c}{$-0.0727$} & 
	\multicolumn{2}{c}{$+0.6857$} & 
	\multicolumn{2}{c}{$+0.9184$} \\                 \hline
 $m_{\tilde{\chi}^0_1}$, $\eta_{\tilde{\chi}^0_1}$ & 
	\multicolumn{2}{c}{40.0 GeV, $+1$} & 
	\multicolumn{2}{c}{54.1 GeV, $+1$} & 
	\multicolumn{2}{c}{67.3 GeV, $+1$} \\
 $m_{\tilde{\chi}^0_2}$, $\eta_{\tilde{\chi}^0_2}$ & 
	\multicolumn{2}{c}{88.2 GeV, $+1$} & 
	\multicolumn{2}{c}{113.2 GeV, $+1$} & 
	\multicolumn{2}{c}{99.7 GeV, $-1$} \\
 State of the $\tilde{\chi}^0_1$& 
	\multicolumn{2}{c}{$(-0.95,0.30,0.08,0.08)$} &
	\multicolumn{2}{c}{$(-0.48,0.66,-0.51,-0.27)$} &
	\multicolumn{2}{c}{$(-0.13,0.19,-0.20,0.95)$} \\ 
\end{tabular}
\end{table}

\end{document}